\documentclass[9pt]{article}
\usepackage{spconf}

\usepackage{cite}
\usepackage{amsmath,amssymb,amsfonts}
\usepackage{algorithmic}
\usepackage{graphicx}
\usepackage{textcomp}
\usepackage{xcolor}
\usepackage[dvipsnames]{xcolor}
\def\BibTeX{{\rm B\kern-.05em{\sc i\kern-.025em b}\kern-.08em
    T\kern-.1667em\lower.7ex\hbox{E}\kern-.125emX}}
    
\usepackage{graphicx}
\usepackage{pgfplots}
\usepackage{subcaption}  
\usepackage{float}
\usepackage{tikz}
\usepackage{pgfplotstable}
\usepackage{changepage}
\usepackage{caption}
\usepackage{bm}
\usepgfplotslibrary{groupplots}
\usepackage{stfloats}
\usepackage{enumitem}

\usepackage[shortcuts, acronym, toc, nonumberlist=false]{glossaries}
\makenoidxglossaries
\newacronym{3gpp}{3GPP}{3rd Generation Partnership Project}
\newacronym{ar}{AR}{Autoregressive}
\newacronym{bic}{BIC}{Bayesian Information Criterion}
\newacronym{em}{EM}{Expectation-Maximization}
\newacronym{gmm}{GMM}{Gaussian Mixture Model}
\newacronym{gs}{GS}{Gohberg-Semencul}
\newacronym{lmmse}{LMMSE}{Linear Minimum Mean Squared Error}
\newacronym{ls}{LS}{least squares}
\newacronym{mimo}{MIMO}{Multiple-Input Multiple-Output}
\newacronym{nmse}{NMSE}{Normalized Mean Square Error}
\newacronym{snr}{SNR}{Signal-to-Noise Ratio}
\newacronym{wss}{WSS}{Wide-Sense Stationary}

\DeclareMathOperator*{\argmax}{\mathrm{argmax}}

\DeclareMathOperator*{\tr}{\mathrm{tr}}
\DeclareMathOperator*{\E}{\mathbb{E}}
\DeclareMathOperator*{\R}{\mathbb{R}}
\DeclareMathOperator*{\C}{\mathbb{C}}

\makeatletter
\renewcommand\section{\@startsection{section}{1}{0pt}%
  {1ex}
  {0.6ex}
  {\bfseries\centering}}
\renewcommand\subsection{\@startsection{subsection}{2}{0pt}%
  {0.8ex}
  {0.4ex}
  {\bfseries}}
\makeatother

\usepackage{fancyhdr}
\fancypagestyle{cfooter}{
  \fancyhf{}

  \cfoot{\scriptsize 
  \copyright 2026 IEEE. Published in the Proceedings of the 2026 IEEE International Conference on Acoustics, Speech and Signal Processing (ICASSP 2026).\\ DOI: 10.1109/ICASSP55912.2026.11462148.}
}

\begin{document}
\pagestyle{cfooter}

\title{Autoregressive-Gaussian Mixture Models: Efficient Generative Modeling of WSS Signals}

\name{Kathrin Klein, Benedikt B\"ock, Nurettin Turan, Wolfgang Utschick}
\address{TUM School of Computation, Information and Technology, Technical University of Munich, Germany\\ Email: \{kathrin.klein; benedikt.boeck; nurettin.turan; utschick\}@tum.de}

\maketitle

\setlength{\abovedisplayskip}{2pt}
\setlength{\belowdisplayskip}{2pt}
\setlength{\abovedisplayshortskip}{0pt}
\setlength{\belowdisplayshortskip}{0pt}

\begin{abstract}
This work addresses the challenge of making generative models suitable for resource-constrained environments like mobile wireless communication systems.
We propose a generative model that integrates \gls{ar} parameterization into a \gls{gmm} for modeling \gls{wss} processes.
By exploiting model-based insights allowing for structural constraints, the approach significantly reduces parameters while maintaining high modeling accuracy.
Channel estimation experiments show that the model can outperform standard \glspl{gmm} and variants using Toeplitz or circulant covariances, particularly with small sample sizes.
For larger datasets, it matches the performance of conventional methods while improving computational efficiency and reducing the memory requirements.
\end{abstract}


\vspace{0.25cm}
\section{Introduction and Motivation}
Generative modeling, the process of learning data distributions from a given dataset, has become a core area in machine learning, with applications in image generation, signal processing, and beyond.
However, state-of-the-art models like Stable Diffusion \cite{rombach2022} are resource-intensive, making them unsuitable for resource-constrained platforms like mobile devices.
This motivates the development of efficient models that retain performance while reducing complexity \cite{turan2024}.

A promising strategy is to incorporate structural assumptions about the data.
One such assumption is modeling the signal to be \gls{wss}, for which the process mean and autocovariance is shift-invariant.
This structure appears in many domains, notably wireless communications, particularly \gls{mimo} systems with uniform linear arrays, and leads to Toeplitz structured covariance matrices enabling efficient parameterizations \cite{fesl2022}.
In addition, \gls{ar} processes capture the covariance structure of \gls{wss} processes well.
By representing a process as a linear function of past values, \gls{ar} models encode temporal or spatial dependencies with only few parameters \cite[Sec. 6-7]{kay1988}.
Importantly, the \gls{ar} parameters implicitly define the inverse of the process's covariance matrix, offering a compact and structured way to model dependencies \cite{boeck2025}.
This property enables their integration into models like \glspl{gmm}, which are tools for generative modeling, that can be adapted to incorporate \gls{ar}-based covariance representations.
Typically trained via the \gls{em} algorithm, \glspl{gmm} model the data distribution as a mixture of Gaussian components, each with its own mean and covariance \cite[Sec. 9.4]{bishop2006}.

We introduce a compact generative model that integrates \gls{ar}-based covariance parameterization into a \gls{gmm} framework, termed \gls{ar}-\gls{gmm}, to model stationary, non-Gaussian data.
The main contributions are:
\begin{itemize}[itemsep=1pt, topsep=2pt, leftmargin=*, parsep=0pt]
    \item The \gls{ar}-\gls{gmm} is lightweight while preserving modeling performance with structured covariances, to make it well-suited for resource-constrained environments.
    \item An \gls{em} algorithm is developed for efficient learning, with applications in wireless channel estimation, and potential in other domains which underlie on \gls{wss} processes.
\end{itemize}

\section{Preliminaries on GMMs}
\glspl{gmm} are probabilistic models that represent distributions utilizing a latent variable.
It models a distribution $p(\bm{x}) $ as a weighted sum of $K$ Gaussian components:
\begin{equation}
    p(\bm{x}) = \sum_{k=1}^{K} \pi_k \mathcal{N}_{\C}(\bm{x}|\bm{\mu}_k, \bm{C}_k),
\end{equation}
where each component $k$ has the mixture weight $\pi_k$, a mean $\bm{\mu}_k$ and a covariance matrix $\bm{C}_k$.
The latent variable $k$ denotes the component responsible for generating $\bm{x}$ \cite[Sec. 9.2]{bishop2006}.

Parameter estimation is performed using the \gls{em} algorithm, which iteratively maximizes the data log-likelihood in the presence of latent variables.
From now on, we define the training dataset as $\bm{X} = \{\bm{x}^{(n)}\}_{n=1}^N$.
At iteration $t$ of the E-step, the posterior probability (responsibility) of $k$, given the generated sample $\bm{x}^{(n)}$ is computed as:
\begin{equation}
    p_t(k | \bm{x}^{(n)}) = \frac{\pi_k \mathcal{N}_{\C}(\bm{x}^{(n)}|\bm{\mu}_k, \bm{C}_k)}{\sum_{j=1}^{K} \pi_j \mathcal{N}_{\C}(\bm{x}^{(n)}|\bm{\mu}_j, \bm{C}_j)}.
\end{equation}
Using the computed responsibilities, in the M-step, the parameters  $\{\pi_k, \bm{\mu}_k, \bm{C}_k\}$ for each component $k$ are updated to maximize the expected complete-data log-likelihood.
The algorithm alternates between E- and M-step until convergence of the log-likelihood or the parameters \cite[Sec. 9.2]{bishop2006}.

\section{Model Derivation: GMMs with AR-Parameterized Covariance Estimation}
Assuming the data is generated by a \gls{wss} process, with shift-invariant mean and autocovariance, the model exploits the resulting Toeplitz structure of the covariance matrix to reduce complexity.
Instead of directly optimizing the means and covariances in a \gls{gmm}, we reparameterize them using \gls{ar} coefficients.
Let $N$ be the number of samples, $M$ the data dimensionality, and $K$ the number of components.
Component-specific parameters are indexed by $k$.
The \gls{ar} order $\omega(k)$ varies per component and is simplified as $\omega$ and conditioned parameters are marked with the subscript $c$ (i.e., $\geq \omega | < \omega$).
We define the reduced dimension $D = M - \omega$.

\subsection{Conditioned Log-Likelihood Estimation}
While the maximization of the unconditional Gaussian log-likelihood of \gls{ar} processes is generally non-convex and challenging to optimize, using the conditional log-likelihood $p(\bm{x}_{\geq \omega}, k | \bm{x}_{< \omega})$ makes this a convex optimization problem.
Therefore, conditioning on the first $\omega$ entries of each sample $\bm{x}^{(n)}$, simplifies the optimization problem significantly.
This allows the optimum to be expressed in closed form \cite{boeck2025}.
Given that we are using a conditional log-likelihood estimation:
\begin{equation}
    p(\bm{x}_{\geq \omega} | \bm{x}_{< \omega}) = \sum_{k=1}^{K} \pi_{k} \mathcal{N}_{\C}(\bm{x}_{\geq \omega} | \bm{x}_{< \omega}; \bm{\mu}_{c,k}, \bm{C}_{c,k}),
\end{equation}
where component $k$ has a mixture weight $\pi_{k} = p(k) \in \R$ and $\mathcal{N}_{\C}(\bm{x}_{\geq \omega} |\bm{x}_{< \omega}; \bm{\mu}_{c,k}, \bm{C}_{c,k}) $ is the conditional Gaussian distribution with mean $\bm{\mu}_{c,k} \in \C^{D}$ and covariance matrix $\bm{C}_{c,k} \in \C^{D \times D}$ for $k$.
Note that we assume $p(k) = \pi_k = \pi_{c,k} = p(k \mid \bm{x}_{< \omega})$, meaning that the first \(\omega\) entries of \(\bm{x}\) provide no information about each $k$.
This assumption simplifies the model without loss of generality, as it merely fixes the prior of the component $k$ early in the sequence.

Next, we will examine the E-step and M-step that we iterate through, along with the adjustments made during each step.
In the E-step, we calculate the posterior probabilities $p_t(k | \bm{x}^{(n)}_{\geq \omega}, \bm{x}^{(n)}_{< \omega})$, at iteration $t$, for each component $k$, which will be called $\gamma_{t,k}^{(n)}$ for more readability in the following:
\begin{equation}\label{eq:responsibilities}
\begin{aligned}
    \gamma_{t,k}^{(n)} :=
    \frac{
        \pi_k 
        \mathcal{N}_{\C}(\bm{x}^{(n)}_{\geq \omega} | \bm{x}^{(n)}_{< \omega};
        \bm{\mu}_{c,k}, \bm{C}_{c,k})
    }{
        \sum_{j=1}^{K} 
        \pi_j 
        \mathcal{N}_{\C}(\bm{x}^{(n)}_{\geq \omega} | \bm{x}^{(n)}_{< \omega}; 
        \bm{\mu}_{c,j}, \bm{C}_{c,j})
    }
\end{aligned}.
\end{equation}
Here, the key distinction lies in the nature of the conditional Gaussian distributions $\mathcal{N}_{\C}(\bm{x}_{\geq \omega} | \bm{x}_{< \omega}; \bm{\mu}_{c,k}, \bm{C}_{c,k})$.
Because $\bm{x}_{\geq \omega}$ and $\bm{x}_{< \omega}$ are jointly Gaussian given $k$, the means $\bm{\mu}_{c,k}$ and covariances $\bm{C}_{c,k}$ of these conditional distributions can be computed in closed form using the standard formulas of the \gls{lmmse} estimator \cite[Sec. 6.5.1]{kay1988}.
This property simplifies the E-step and also emphasizes that these updates are specific to the conditional framework, differentiating them from an unconditional E-step.
Using the responsibilities in \eqref{eq:responsibilities}, the M-step involves maximizing the expected conditional log-likelihood of the complete data with respect to $\{\pi_k\}_{k=1}^K, \{\bm{\mu}_{c,k}\}_{k=1}^K$ and $\{\bm{C}_{c,k}\}_{k=1}^K$, which will be given as $ \bm{\pi}, \bm{\mu} \text{ and } \bm{C}$ \cite[Sec. 9.2]{bishop2006}:
\begin{equation}
    \argmax_{\bm{\pi}, \bm{\mu}, \bm{C}} 
    \sum_{n=1}^{N} \mathbb{E}_{\gamma_{t,k}^{(n)}} 
    \Big[
        \ln p(\bm{x}^{(n)}_{\geq \omega} | k, \bm{x}^{(n)}_{< \omega}) + \ln p(k)
    \Big],
\end{equation}
where the first term captures the log-likelihood of the data given component $k$ and the second term represents the prior probability of component $k$, by assuming $p(k) = p(k \mid \bm{x}_{< \omega})$.
We now focus on the data log-likelihood term $p(\bm{x}^{(n)}_{\geq \omega} | k, \bm{x}^{(n)}_{< \omega})$, which follows a Gaussian distribution.
For a shorter representation, we use the identity $\tr(\bm{A}\bm{B}\bm{C}) = \tr(\bm{C}\bm{A}\bm{B})$ \cite[Sec. 0.2.5]{horn2013} and introduce the sample covariance matrix $\bm{S}^{(n)}_{\geq \omega}= (\bm{x}^{(n)}_{\geq \omega} - \bm{\mu}_{c,k}) (\bm{x}^{(n)}_{\geq \omega} - \bm{\mu}_{c,k})^{\operatorname{H}}$:
\begin{equation}\label{eq:conditional-gmm-result}
\begin{aligned}
    &\sum_{n=1}^{N} \mathbb{E}_{\gamma_{t,k}^{(n)}} \Big[ \ln p(\bm{x}^{(n)}_{\geq \omega} | k, \bm{x}^{(n)}_{< \omega}) \Big]
    = - \sum_{n=1}^{N} \sum_{k=1}^{K} 
        \gamma_{t,k}^{(n)} \\
        &\Big[
            D \ln(\pi)
            + \ln \det \bm{C}_{c,k}
            + \tr\left( \bm{C}_{c,k}^{-1} 
            \bm{S}^{(n)}_{\geq \omega} \right)
        \Big].
\end{aligned}
\end{equation}
In the following, we assume zero-mean data, a practical simplification that shifts focus entirely to modeling covariance.
While $\E[\bm{x}] = 0$ holds generally, with $\bm{x}$ being the channel state vector, this doesn't imply $\E[\bm{x} | k] = 0$ for all $k$, a nuance relevant in wireless applications \cite{boeck2024}.
Therefore, we still need to fit the conditional means $\bm{\mu}_{c,k}^{(n)}$ for each component $k$ given by $\bm{\mu}_{c,k}^{(n)} = \bm{C }_{(\geq \omega,< \omega),k} \bm{C}_{< \omega,k}^{-1} \bm{x}_{< \omega}^{(n)}$.
The conditional covariance matrix $\bm{C}_{c,k}$ can be determined as $\bm{C}_{c,k} = \bm{C}_{\geq \omega,k} - \bm{C}_{(\geq \omega,<\omega),k} \bm{C}_{<\omega,k}^{-1} \bm{C}_{(<\omega, \geq \omega),k}$ \cite[Sec. 2.3.1]{bishop2006}, where
$\bm{C}_k = \begin{bmatrix} 
    \bm{C}_{<\omega,k} & \bm{C}_{(<\omega,\geq\omega),k}\\
    \bm{C}_{(\geq\omega,<\omega),k} & \bm{C}_{\geq\omega,k} 
\end{bmatrix}$. 
From this relation between $\bm{C}_{c,k}$ and $\bm{C}_{k}$ it also becomes clear that by optimizing the conditional log-likelihood with $\bm{C}_{c,k}$, we are implicitly fitting an unconditional \gls{gmm} using $\bm{C}_{k}$ to the data.

\subsection{Introduction to AR Parameters}
An \gls{ar} model of order $ \omega $ is defined as $x_i = \sum_{j=1}^{\omega} a_j x_{i-j} + \epsilon_i, \quad \epsilon_i \sim \mathcal{N}_{\C}(0, \sigma^2)$
where $x_i$ is the current value, $\{a_j\}_{j=1}^{\omega} $ are the \gls{ar} coefficients, and $ \epsilon_i $ is Gaussian noise with zero mean and variance $ \sigma^2 $ \cite[Sec. 2.1.3]{kirchgassner2007}.
The parameter $\omega$ donates the order of the \gls{ar} model, indicating how many past values influence the current value.
Substituting the log-likelihood of the observations $ x_\omega, \dots, x_{M-1} $ given the past data $ x_0, \dots, x_{\omega-1} $ into \eqref{eq:conditional-gmm-result}, we get the following form of the log-likelihood to maximize \cite[Sec. 6.5.1]{kay1988}:
\begin{equation}\label{eq:conditional-gmm-result-ar}
\begin{aligned}
    \argmax_{\bm{a}, \bm{\sigma^2}} \sum_{n=1}^{N} \sum_{k=1}^{K} \gamma_{t,k}^{(n)}
    \left[ -D \ln(\pi \sigma_k^2) - \frac{1}{\sigma_k^2} \|\bm{x}^{(n)}_{\geq \omega} - \bm{\mu}^{(n)}_{c}(\bm{a}_k)\|^2_2 \right]
\end{aligned}
\end{equation}
Here we use $\bm{a} = \{\bm{a}_k\}_{k=1}^K$ and $\bm{\sigma^2} = \{\sigma^2_k\}_{k=1}^K$, where $ \sigma_k^2 $ represents the variance for the $k$-th component, and $ \bm{a}_k \in \C^{\omega} $ is the vector of \gls{ar} parameters for component $k$.
The notation $ \bm{\mu}^{(n)}_{c}(\bm{a}_k) $ refers to the model's predicted value for the $n$-th data point, the $i$-th entry is given by
$ \sum_{m=1}^{\omega} a_m x_{i - m} $, where $ x_{i - m} $ denotes the $(i - m)$-th entry of the vector $\bm{x}$.
To optimize the objective with respect to the \gls{ar} parameters $\bm{a}_{\tilde{k}} \in \R^{\omega}$, for a specific $\tilde{k}$, we define the error term as:
\begin{equation}
    E(\bm{a}_{\tilde{k}}) =  \sum_{n=1}^{N} \gamma_{t,\tilde{k}}^{(n)} \frac{1}{\sigma_{\tilde{k}}^2} \left[ \|\bm{x}^{(n)} _{\geq \omega} - \bm{A}^{(n)} \bm{a}_{\tilde{k}}\|_2^2 \right],
\end{equation}
where $ \bm{A}^{(n)} $ is the matrix of auto-regressive terms for the $n$-th data point, with $A^{(n)}_{i,j} = x^{(n)}_{i + \omega - 1 - j}, \quad \text{for } i = 0, \dots, D,\; j = 0, \dots, \omega - 1$.
By solving a normal equation, we obtain:
\begin{equation}
    \bm{a}_{\tilde{k}}^{\text{new}} = \left( \sum_{n=1}^{N} \gamma_{t,\tilde{k}}^{(n)} \bm{A}^{(n), \operatorname{H}} \bm{A}^{(n)} \right)^{-1} \left( \sum_{n=1}^{N} \gamma_{t,\tilde{k}}^{(n)} \bm{A}^{(n),\operatorname{H}} \bm{x}^{(n)}_{\geq \omega} \right).
\end{equation}
To maximize the log-likelihood with respect to $\sigma_{\tilde{k}}^2$, we differentiate the log-likelihood function in \eqref{eq:conditional-gmm-result-ar} with respect to $\sigma_{\tilde{k}}^2$:
\begin{equation}
    \sigma_{\tilde{k}}^{2, \text{new}} = \frac{\sum_{n=1}^{N} \gamma_{t,\tilde{k}}^{(n)} \|\bm{x}^{(n)}_{\geq \omega} - \bm{\mu}^{(n)}_{c}(\bm{a}_{\tilde{k}}^{\text{new}})\|_2^2}{D\sum_{n=1}^{N} \gamma_{t,\tilde{k}}^{(n)}}.
\end{equation}
Using a Lagrangian with constraint $\sum_{k=1}^{K} \pi_k = 1$, we obtain:
\begin{equation}
    \pi_{\tilde{k}}^{\text{new}} = \frac{N_{\tilde{k}}}{N}, \quad \text{where} \quad N_{\tilde{k}} = \sum_{n=1}^{N} \gamma_{t,\tilde{k}}^{(n)}.
\end{equation}

\subsection{Gohberg-Semencul Estimation for Conditional Mean and Covariance with Positive-Definiteness Constraints}
To compute the conditional mean $\bm{\mu}_{c,k}^{\text{new}}$ and covariance $\bm{C}_{c,k}^{\text{new}}$ for the next E-step, we use the updated parameters $\bm{a}_k^{\text{new}}$ and $\sigma_k^{2, \text{new}}$.
While the \gls{ar} reparameterization provides a compact and structure-preserving description of the covariance matrix, we must ensure that the reconstructed matrices still are positive definite to guarantee a stable \gls{em}-algorithm.
This is achieved by imposing box constraints on \gls{gs} coefficients:
\begin{equation}
    \left | \frac{\alpha_i}{\alpha_0} \right | \leq K_i, \quad K_i > 0, \quad i = 1, \dots, \omega-1,
\end{equation}
which are closely related to the \gls{ar} parameters according to $\alpha_{0,k} = 1/\sigma_k^{2,\text{new}}$ and $\alpha_{i,k} = -a_{i,k}^{\text{new}} / \sigma_k^{2,\text{new}}$ \cite{boeck2025, kailath1975} and, thus, imposing box constraints for the \gls{gs} coefficients implies box constraints for the corresponding \gls{ar} parameters. 
These constraints ensure that the \gls{gs}-based inverse covariance matrix $\bm{\Gamma}_k$  lies in a class of matrices guaranteed to be invertible and stable \cite{boeck2025, mukherjee1988}.
$\bm{\Gamma}_k $ is given as:
\begin{equation}
    \bm{C}_k^{-1} = \bm{\Gamma}_k = \frac{1}{\alpha_0} (\bm{B}_k\bm{B}_k^{\operatorname{H}} - \bm{Z}_k\bm{Z}_k^{\operatorname{H}}).
\end{equation}
Matrices $\bm{B}_k$ and $\bm{Z}_k$ are lower-triangular Toeplitz matrices built from the entities of $\alpha$, as defined in \cite{boeck2025, mukherjee1988}.

With these additional constraint, \eqref{eq:conditional-gmm-result-ar} reads as:
\begin{equation}
\begin{aligned}
    \argmax_{\bm{a}, \sigma^2} \; & \sum_{n=1}^{N} \sum_{k=1}^{K} \gamma_{t,k}^{(n)}
    \left[ -D \ln(\pi \sigma_k^2) 
    - \frac{1}{\sigma_k^2} 
    \| \bm{x}^{(n)}_{\geq \omega} - \bm{\mu}^{(n)}_{c}(\bm{a}_k) \|_2^2 \right] \\
    \text{s.t.} \quad & |a_i| \leq K_i, \quad i = 1, \dots, \omega-1.
\end{aligned}
\end{equation}
To solve this, we apply direct projection of the update step onto the constraint set defined by $|a_i| \leq K_i$.
Where $K_i$ are fixed over the training iterations, which use an exponential decay with factor $\lambda_k$ for each $k$.
Our simulations indicate that directly projecting $a_i$ onto the constraint $K_i$ yields favorable results.
Consequently, we present results obtained using this direct projection approach, as it is computationally more efficient than projected gradient descent and avoids adding complexity to the algorithm, aligning with our objective of maintaining computational efficiency.

\section{Parameter Count}
In a standard \gls{gmm} with complex data, full covariance matrices and zero-means, each of the $K$ components in a $M$-dimensional space requires $M^2$ parameters for its Hermitian covariance matrix, plus $K - 1$ mixing coefficients \cite[Sec. 9.2]{bishop2006}, resulting in a total parameter count of $K \cdot M^2 + (K - 1)$.
Variants that exploit a Toeplitz or circulant covariance matrix structure, as introduced in \cite{fesl2022}, already reduce the parameter count (see Table \ref{tab:param_count_comparison}).
In contrast, our proposed model incorporates \gls{ar} coefficients, yielding an even greater reduction in complexity.
Each component uses $2\omega_k$ \gls{ar} parameters, one noise variance, and there are overall  $K - 1$ mixing coefficients.
The total count is $\sum_{k=1}^{K} 2\omega_k + 2K - 1$.
As shown in Table~\ref{tab:param_count_comparison}, for $K = 16$ and $M = 64$, the standard \gls{gmm} requires 65,551 parameters, whereas the proposed model, using $\omega_k = 2$, chosen based on the observed average value around 1--2 from hyperparameter tuning, requires only 95 parameters.
This reduction, by a factor $\rho_{\mathrm{red}} = 690$, highlights the model's memory efficiency, especially in high-dimensional settings.
\begin{table}[!t]
\centering
\caption{Comparison of parameter counts for standard, structured and proposed \gls{gmm}, all assuming zero means, with example ($K = 16$, $M = 64$, $\omega_k = 2$) and reduction factor $\rho_{\mathrm{red}}$.}
\label{tab:param_count_comparison}
\vspace{-0.25cm}
\begin{tabular}{|l|c|r|r|}
\hline
\textbf{Model} & \textbf{Number of Parameters} & \textbf{Example} & \textbf{$\rho_{\mathrm{red}}$} \\\hline
Std. \gls{gmm} & $K \cdot M^2 + (K - 1)$ & $65,551$ & $1$ \\ \hline
Toeplitz & $K \cdot 2M + (K - 1)$ & $2,063$ & $31.8$ \\ \hline
Circulant & $K \cdot M + (K - 1)$ & $1,039$ & $63$ \\ \hline
\gls{ar}-\gls{gmm} & $\sum_{k=1}^{K} 2 \omega_k + 2K - 1$ & $95$ & $690$ \\ \hline
\end{tabular}
\vspace{-0.5cm}
\end{table}

\section{Application to 3GPP Channel Data}
We apply the proposed \gls{ar}-\gls{gmm} to synthetic \gls{3gpp} channel data generated using the model \cite{neumann2018}, which is based on \cite{3gpp2020}.
Generation depends on the number of samples $N$, the number of antennas $M$, and the number of propagation clusters $P$.
The generated channel samples are used for training, and the resulting \gls{gmm} is utlized for channel estimation.
The channel estimates are based on the observed data and the learned posterior probabilities of the \gls{gmm}, by weighting the conditional expectations given by the \gls{lmmse} formula with these posterior probabilities, as described in detail in \cite{koller2022}.
The estimator's performance is evaluated using the \gls{nmse} over different \gls{snr} values and is compared against these baselines: a \gls{lmmse} estimator, which estimates covariance directly from the data (\gls{gmm} with only one component), a full-covariance \gls{gmm}, Toeplitz and circulant variants, the \gls{ls} estimator, and a genie-aided \gls{lmmse} estimator with perfect second-order statistics \cite[Sec. 8.5.3, 9.1.6]{hossein2014}.

\subsection{Hyperparameter Tuning}
To optimize the \gls{ar}-\gls{gmm}, we tune the \gls{ar} order $\omega_k$ and the regularization factor $\lambda_k$ per component, which we are using as a factor in an exponential decay strategy for the box constraints.
Optimization is guided by minimizing both \gls{nmse} \cite[Sec. 9.1.5]{hossein2014} and the \gls{bic} \cite[Sec. 9]{sadanori2008}.
Random search explores $10^3$ hyperparameter configurations ($10^2$ for $N \geq 10^4$), followed by fine-tuning with Optuna \cite{akiba2019}, which refines the search around high-performing regions.
We evaluate across multiple settings for different values of $N = \{ 10^2,\ 10^3,\ 10^4 \}$, as well as different dimensions $ M = \{ 16, 32, 64 \}$, to assess the model’s performance across varying training dataset sizes and number of antennas.
Additionally, we test with different numbers of components $ K = \{ 1, 2, 4, 8, 16, 32 \}$, to investigate the impact of the number of components on the model’s performance.
Furthermore, we incorporate different values for the number of paths, considering $P = \{1, 3\}$, to study the effect of multipath propagation.
Across all experiments, optimal performance typically occurs at very low \gls{ar} orders (often 1–2), indicating strong robustness and limited sensitivity to model-order mismatch.

\section{Simulation Results and Analysis}
\begin{figure*}[!t]
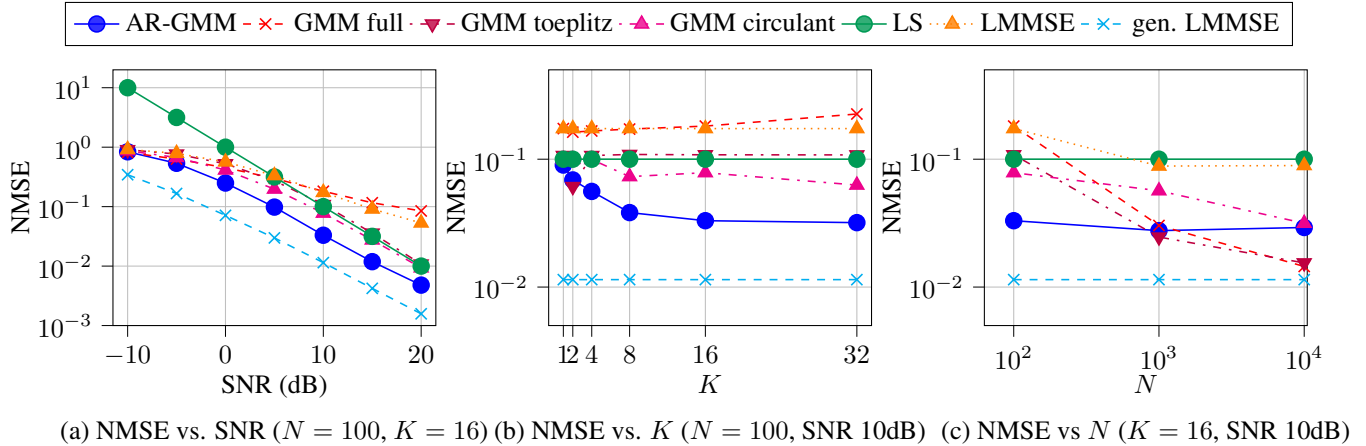
 
    \centering
    \hspace{0.1cm}
    \vspace{0.15cm}
    \ref{sharedLegend}
    \begin{tikzpicture}
        \begin{groupplot}[
            group style={
                group size=3 by 1,
                horizontal sep=1.5cm,
            },
            width=0.33\textwidth,
            height=0.28\textwidth,
            legend to name=sharedLegend,
            legend columns=7,
            legend style={fill opacity=0.8, draw opacity=1, text opacity=1, anchor=south},
        ]
        \input{images/NMSE_SNR_Performance_N100_M64_K16_paths1}
        \input{images/NMSE_K_Performance_N100_M64_paths1_SNR10}
        \input{images/NMSE_N_Performance_M64_K16_paths1_SNR10}
        \end{groupplot}
    \end{tikzpicture}
    \vspace{-0.75cm}
    \caption{\gls{nmse} over (a) different SNRs, (b) the number of components $K$ and (c) training samples sizes $N$ ($M=64$, $P=1$).}
    \label{fig:NMSE-plots}
    \vspace{-0.5cm}
\end{figure*}
This section presents simulation results for the proposed \gls{ar}-\gls{gmm} applied to the \gls{3gpp} channel data, which gives the ground truth.
We evaluate \gls{nmse} performance across different \gls{snr} values for $N = 10^2$, $M = 64$, $K = 16$, and $P = 1$,  see Fig. \ref{fig:NMSE-plots} (a).
It shows that our \gls{ar}-\gls{gmm} outperforms standard and structured \gls{gmm}s, especially at low sample sizes and high dimensions.
Additionally, the \gls{ls} estimator, is shown along with the genie-aided \gls{lmmse} estimator as a lower and upper bound for the performance.
Although the performance gap narrows with larger datasets, our model maintains comparable accuracy with far fewer parameters.
Fig. \ref{fig:NMSE-plots} (b) illustrate how model performance varies with the number of components $K$, for $N = 10^2$, $M = 64$, and $P = 1$, at \gls{snr} 10dB.
Increasing $K$ generally improves \gls{nmse}, with diminishing returns beyond $K=16$ or $K=32$, indicating a saturation effect.
While our \gls{ar}-\gls{gmm} continues to perform robustly, the full covariance \gls{gmm} begins to overfit, especially at small $N$ and is even worse then \gls{ls}.
The genie-aided \gls{lmmse} provides a lower bound on the \gls{nmse}, i.e., an upper bound on achievable performance.
Overall, these results confirm that our model achieves strong performance with moderate $K$, offering an efficient trade-off between accuracy and complexity.
Fig. \ref{fig:NMSE-plots} (c) shows \gls{nmse} versus training set size $N$ for $M = 64$, $K = 16$, and \gls{snr} = 10dB with a single path $P = 1$.
The plot reveals that our \gls{ar}-\gls{gmm} outperforms benchmarks at low $N$, particularly around $N = 10^2$.
However, with increasing training dataset size, \gls{gmm} full and \gls{gmm} Toeplitz outperform the \gls{ar}-\gls{gmm}.
In consequence, the proposed \gls{ar}-\gls{gmm} is particularly well suited for situations with only few training samples or situations, in which the model parameter-count is critical, e.g., when offloading the model to user equipments \cite{turan2024}.

Overall, the results demonstrate that our \gls{ar}-\gls{gmm} excels in scenarios with small training sizes and a high number of antennas.
Additional configurations, multipath cases and alternative strategies using Optuna yielded similar trends.

\section{Conclusion and Outlook}
This work introduced a novel generative model for \gls{wss} processes by embedding \gls{ar} parameterization into a \gls{gmm}, leveraging the Toeplitz covariance structure to reduce model complexity.
We demonstrate that the proposed model can be used for channel estimation, where it is especially effective in small-sample scenarios, outperforming common \gls{gmm} variants.
Overall, the approach offers an efficient alternative to traditional models, with reduced parameter requirements.

Future work could validate the model on real-world data and investigate how performance evolves under approximate \gls{wss} or Toeplitz conditions.
Beyond wireless communications, this framework may also benefit domains like audio and finance, where modeling stationary dependencies is key.

\bibliographystyle{IEEEbib}
\bibliography{IEEEabrv,references}

@book{bishop2006,
  title={{Pattern Recognition and Machine Learning}},
  author={C. M. Bishop},
  year={2006},
  publisher={Springer Science+Business Media, LLC}
}

@article{boeck2025,
  author={B. Böck and D. Semmler and B. Fesl and M. Baur and W. Utschick},
  journal={IEEE Transactions on Signal Processing}, 
  title={{Gohberg-Semencul Toeplitz Covariance Estimation via Autoregressive Parameters}}, 
  year={2025},
  volume={73},
  number={},
  pages={858-875}
}

@article{koller2022,
  author={M. Koller and B. Fesl and N. Turan and W. Utschick},
  journal={IEEE Transactions on Signal Processing}, 
  title={{An Asymptotically MSE-Optimal Estimator Based on Gaussian Mixture Models}}, 
  year={2022},
  volume={70},
  pages={4109-4123}}

@article{turan2024,
  author    = {N. Turan and B. Fesl and M. Koller and M. Joham and W. Utschick},
  title     = {{A versatile low-complexity feedback scheme for FDD systems via generative modeling}},
  journal   = {IEEE Transactions on Wireless Communications},
  volume    = {23},
  number    = {6},
  pages     = {6251--6265},
  year      = {2024}
}

@article{boeck2024,
  author={B. Böck and M. Baur and N. Turan and D. Semmler and W. Utschick},
  journal={IEEE Wireless Communications Letters}, 
  title={{A Statistical Characterization of Wireless Channels Conditioned on Side Information}}, 
  year={2024},
  volume={13},
  number={12},
  pages={3508-3512}
}

@article{neumann2018,
  author={D. Neumann and T. Wiese and W. Utschick},
  journal={IEEE Transactions on Signal Processing}, 
  title={{Learning the MMSE Channel Estimator}}, 
  year={2018},
  volume={66},
  number={11},
  pages={2905-2917}
}

@techreport{3gpp2020,
  author    = {3GPP},
  title     = {{Spatial channel model for multiple input multiple output (MIMO) simulations}},
  institution = {3rd Generation Partnership Project (3GPP)},
  type      = {Tech. Rep.},
  number    = {25.996 (V16.0.0)},
  year      = {2020},
  month     = {July}
}

@book{sadanori2008,
  author    = {K. Sadanori and K. Genshiro},
  title     = {{Information Criteria and Statistical Modeling}},
  publisher = {Springer},
  year      = {2008}
}

@book{hossein2014,
  author    = {P. Hossein},
  title     = {{Introduction to Probability, Statistics, and Random Processes}},
  publisher = {Kappa Research LLC},
  year      = {2014},
  note      = {[Online]. Available: https://www.probabilitycourse.com}
}

@inproceedings{akiba2019,
    title={{Optuna: A Next-generation Hyperparameter Optimization Framework}},
    author={T. Akiba and S. Sano and T. Yanase and T. Ohta and M. Koyama},
    booktitle={Proceedings of the 25th {ACM} {SIGKDD} International Conference on Knowledge Discovery and Data Mining},
    year={2019}
}

@article{mukherjee1988,
  author  = {B. Mukherjee and S. Maiti},
  title   = {{On some properties of positive definite Toeplitz matrices and their possible applications}},
  journal = {Linear Algebra and its Applications},
  volume  = {102},
  pages   = {211--240},
  year    = {1988}
}

@book{kirchgassner2007,
  author    = {G. Kirchgässner and J. Wolters},
  title     = {{Introduction to Modern Time Series Analysis}},
  publisher = {Springer Verlag},
  year      = {2007},
  month     = jan
}

@article{kailath1975,
  author  = {T. Kailath and A. Vieira and M. Morf},
  title   = {{Inverses of Toeplitz operators, innovations, and orthogonal polynomials}},
  journal = {Proceedings of the 1975 IEEE Conference on Decision Control including the 14th Symposium on Adaptive Processes},
  year    = {1975},
  pages   = {749--754}
}

@book{horn2013,
  title     = {{Matrix Analysis}},
  author    = {R. Horn and C. Johnson},
  edition   = {2nd},
  year      = {2013},
  publisher = {Cambridge University Press}
}

@book{kay1988,
  author    = {S. Kay},
  title     = {{Modern Spectral Estimation: Theory and Application}},
  series    = {Prentice-Hall Signal Processing Series},
  publisher = {Prentice Hall},
  year      = {1988},
}

@inproceedings{rombach2022,
  author = {R. Rombach and A. Blattmann and D. Lorenz and P. Esser and B. Ommer},
  booktitle={2022 IEEE/CVF Conference on Computer Vision and Pattern Recognition (CVPR)}, 
  title={{High-Resolution Image Synthesis with Latent Diffusion Models}}, 
  year={2022},
  pages={10674-10685}
}

@inproceedings{fesl2022,
  author={B. Fesl and M. Joham and S. Hu and M. Koller and N. Turan and W. Utschick},
  booktitle={2022 56th Asilomar Conference on Signals, Systems, and Computers}, 
  title={{Channel Estimation based on Gaussian Mixture Models with Structured Covariances}}, 
  year={2022},
  pages={533-537}
}

\end{document}